\documentclass[a4paper]{article}

\usepackage{INTERSPEECH2019}
\usepackage{multirow}
\usepackage{hyperref}

\newcommand{\astfootnote}[1]{
\let\oldthefootnote=\thefootnote
\setcounter{footnote}{0}
renewcommand{\thefootnote}{\fnsymbol{footnote}}
\footnote{#1}
\let\thefootnote=\oldthefootnote
}

\newcommand*\samethanks[1][\value{footnote}]{\footnotemark[#1]}

\title{Temporal Convolution for Real-time Keyword Spotting on Mobile Devices}

\name{Seungwoo Choi\sthanks{\quad Equal contributions, listed in alphabetical order.}, Seokjun Seo\samethanks[1], Beomjun Shin\samethanks[1], Hyeongmin Byun, \\
Martin Kersner, Beomsu Kim, Dongyoung Kim\sthanks{\quad Shared corresponding authors.}, Sungjoo Ha\samethanks[2]}

\address{Hyperconnect, Seoul, South Korea}
\email{\{seungwoo.choi, seokjun.seo, beomjun.shin, hyeongmin.byun\}@hpcnt.com \\ \{martin.kersner, beomsu.kim, dongyoung.kim,  shurain\}@hpcnt.com}

\begin{document}
\maketitle

\begin{abstract}
Keyword spotting (KWS) plays a critical role in enabling speech-based user interactions on smart devices.
Recent developments in the field of deep learning have led to wide adoption of convolutional neural networks (CNNs) in KWS systems due to their exceptional accuracy and robustness.
The main challenge faced by KWS systems is the trade-off between high accuracy and low latency.
Unfortunately, there has been little quantitative analysis of the actual latency of KWS models on mobile devices.
This is especially concerning since conventional convolution-based KWS approaches are known to require a large number of operations to attain an adequate level of performance.

In this paper, we propose a temporal convolution for real-time KWS on mobile devices.
Unlike most of the 2D convolution-based KWS approaches that require a deep architecture to fully capture both low- and high-frequency domains, we exploit temporal convolutions with a compact ResNet architecture.
In Google Speech Command Dataset, we achieve more than \textbf{385x} speedup on Google Pixel 1 and surpass the accuracy compared to the state-of-the-art model.
In addition, we release the implementation of the proposed and the baseline models including an end-to-end pipeline for training models and evaluating them on mobile devices.
\end{abstract}

\noindent\textbf{Index Terms}: keyword spotting, real-time, convolutional neural network, temporal convolution, mobile device

\section{Introduction}

Keyword spotting (KWS) aims to detect pre-defined keywords in a stream of audio signals.
It is widely used for hands-free control of mobile applications.
Since its use is commonly concentrated on recognizing wake-up words (e.g., ``Hey Siri''~\cite{sigtia-interspeech-2018-heysiri}, ``Alexa''~\cite{sun-interspeech-2017-alexa, tucker-interspeech-2016-alexa}, and ``Okay Google''~\cite{chen-icassp-2014-OkayGoogle}) or distinguishing common commands (e.g., ``yes'' or ``no'') on mobile devices, the response of KWS should be both \emph{immediate} and \emph{accurate}.
However, it is challenging to implement fast and accurate KWS models that meet the real-time constraint on mobile devices with restricted hardware resources.

Recently, with the success of deep learning in a variety of cognitive tasks, neural network based approaches have become popular for KWS~\cite{wang-arxiv-2017-ctc, sainath-interspeech-2015-cnn,  zhang-arxiv-2017-helloedge, tang-icassp-2018-residual, deandrade-arxiv-2018-attention, arik-interspeech-2017-crnn}.
Especially, KWS studies based on convolutional neural networks (CNNs) show remarkable accuracy~\cite{sainath-interspeech-2015-cnn, zhang-arxiv-2017-helloedge, tang-icassp-2018-residual}.
Most of CNN-based KWS approaches receive features, such as mel-frequency cepstral coefficients (MFCC), as a 2D input of a convolutional network.
Even though such CNN-based KWS approaches offer reliable accuracy, they demand considerable computations to meet a performance requirement.
In addition, inference time on mobile devices has not been analyzed quantitatively, but instead, indirect metrics have been used as a proxy to the latency.
Zhang \emph{et~al}.~\cite{zhang-arxiv-2017-helloedge} presented the total number of multiplications and additions performed by the whole network.
Tang and Lin~\cite{tang-icassp-2018-residual} reported the number of multiplications of their network as a surrogate for inference speed.
Unfortunately, it has been pointed out that the number of operations such as additions and multiplications, is only an indirect alternative for the direct metric such as latency~\cite{sandler-cvpr-2018-mobilenetv2, tan-arxiv-2018-mnasnet, ma-eccv-2018-shufflenetv2}.
Neglecting the memory access costs and different platforms being equipped with varying degrees of optimized operations are potential sources for the discrepancy.
Thus, we focus on the measurement of actual latency on mobile devices.

In this paper, we propose a temporal convolutional neural network for real-time KWS on mobile devices, denoted as \emph{TC-ResNet}.
We apply \emph{temporal convolution}, i.e., 1D convolution along the temporal dimension, and treat MFCC as input channels.
The proposed model utilizes advantages of temporal convolution to enhance the accuracy and reduce the latency of mobile models for KWS.
Our contributions are as follows:

\begin{itemize}
    \item We propose \emph{TC-ResNet} which is a \emph{fast} and \emph{accurate} convolutional neural network for real-time KWS on mobile devices. According to our experiments on Google Pixel 1, the proposed model shows \textbf{385x} speedup and a 0.3\%p increase in accuracy compared to the state-of-the-art CNN-based KWS model on Google Speech Commands Dataset~\cite{googlespeechcommandsv1}.
    \item We release our models\footnote{Source code can be found at the following link: \url{https://github.com/hyperconnect/TC-ResNet}} for KWS and implementations of the state-of-the-art CNN-based KWS models~\cite{sainath-interspeech-2015-cnn, zhang-arxiv-2017-helloedge, tang-icassp-2018-residual} together with the complete benchmark tool to evaluate the models on mobile devices.
    \item We empirically demonstrate that temporal convolution is indeed responsible for reduced computation and increased performance in terms of accuracy compared to 2D convolutions in KWS on mobile devices.
\end{itemize}

\section{Network Architecture}\label{section:architecture}

\subsection{Temporal Convolution for KWS}\label{subsection:1dconv}
\begin{figure}[h]
    \includegraphics[width=.47\textwidth]{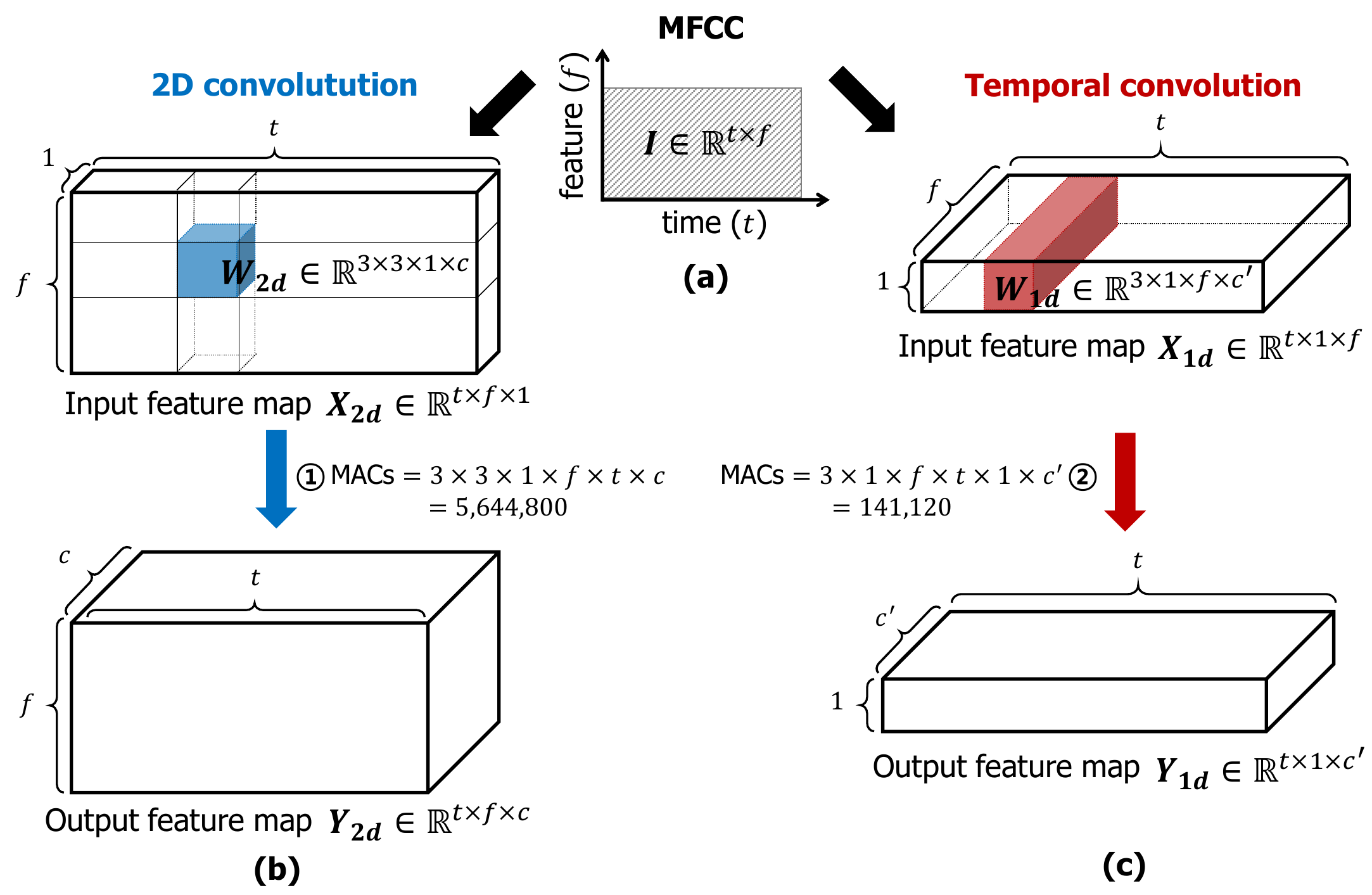}
    \caption{A simplified example illustrating the difference between 2D convolution and temporal convolution. (a) MFCC. (b) 2D convolution for conventional CNN-based KWS approaches. (c) Proposed temporal convolution. Note that both the parameters of a conventional 2D convolution and that of the temporal convolution have the same size in this example by setting $t=98$, $f=40$, $c=160$, and $c^{\prime}=12$.}
    \vspace{-0.2cm}
    \label{fig:1dconv}
\end{figure}

Figure~\ref{fig:1dconv} is a simplified example illustrating the difference between 2D convolution and temporal convolution for KWS approaches utilizing MFCC as input data.
Assuming that stride is one and zero padding is applied to match the input and the output resolution, given input $\mathbf{X} \in \mathbb{R}^{w \times h \times c}$ and weight $\mathbf{W} \in \mathbb{R}^{k_{w} \times k_{h} \times c \times c^{\prime}}$, 2D convolution outputs $\mathbf{Y} \in \mathbb{R}^{w \times h \times c^{\prime}}$.
MFCC is widely used for transforming raw audio into a time-frequency representation, $\mathbf{I} \in \mathbb{R}^{t \times f}$, where $t$ represents the time axis ($x$-axis in Figure~\ref{fig:1dconv}a) and $f$ denotes the feature axis extracted from frequency domain ($y$-axis in Figure~\ref{fig:1dconv}a).
Most of the previous studies~\cite{zhang-arxiv-2017-helloedge, tang-icassp-2018-residual} use input tensor $\mathbf{X} \in \mathbb{R}^{w \times h \times c}$ where $w = t$, $h = f$ (or vice versa), and $c = 1$ ($\mathbf{X_{2d}} \in \mathbb{R}^{t \times f\times 1}$ in Figure~\ref{fig:1dconv}b).

CNNs are known to perform a successive transformation of low-level features into higher level concepts.
However, since modern CNNs commonly utilize small kernels, it is difficult to capture informative features from both low and high frequencies with a relatively shallow network (colored box in Figure~\ref{fig:1dconv}b only covers a limited range of frequencies).
Assuming that one naively stacks $n$ convolutional layers of $3 \times 3$ weights with a stride of one, the receptive field of the network only grows up to $2n + 1$.
We can mitigate this problem by increasing the stride or adopting pooling, attention, and recurrent units.
However, many models still require a large number of operations, even if we apply these methods, and has a hard time running real-time on mobile devices.

In order to implement a \emph{fast} and \emph{accurate} model for real-time KWS, we reshape the input from $\mathbf{X_{2d}}$ in Figure~\ref{fig:1dconv}b to $\mathbf{X_{1d}}$ in Figure~\ref{fig:1dconv}c.
Our main idea is to treat per-frame MFCC as a time series data, rather than an intensity or grayscale image, which is a more natural way to interpret audio.
We consider $\mathbf{I}$ as one-dimensional sequential data whose features at each time frame are denoted as $f$.
In other words, rather than transforming $\mathbf{I}$ to $\mathbf{X_{2d}} \in \mathbb{R}^{t \times f\times 1}$, we set $h = 1$ and $c = f$, which results in $\mathbf{X_{1d}} \in \mathbb{R}^{t \times 1\times f}$, and feed it as an input to temporal convolution (Figure~\ref{fig:1dconv}c). 
The advantages of the proposed method are as follows:

\textbf{Large receptive field of audio features.} 
In the proposed method, all lower-level features always participate in forming the higher-level features in the next layer.
Thus, it takes advantage of informative features in lower layers (colored box in Figure~\ref{fig:1dconv}c covers a whole range of frequencies), thereby avoiding stacking many layers to form higher-level features.
This enables us to achieve better performance even with a small number of layers.

\textbf{Small footprint and low computational complexity.} 
Applying the proposed method, a two-dimensional feature map shrinks in size if we keep the number of parameters the same as illustrated in Figure~\ref{fig:1dconv}b and \ref{fig:1dconv}c.
Assuming that both conventional 2D convolution, $\mathbf{W_{2d}} \in \mathbb{R} ^ {3 \times 3 \times 1 \times c}$, and proposed temporal convolution, $\mathbf{W_{1d}} \in \mathbb{R}^{3 \times 1 \times f \times c^{\prime}}$, have the same number of parameters (i.e., $c^{\prime}=\frac{3 \times c}{f}$), the proposed temporal convolution requires a smaller number of computations compared to the 2D convolution (\textcircled{\raisebox{-0.9pt}{2}} is smaller than \textcircled{\raisebox{-0.9pt}{1}} in Figure~\ref{fig:1dconv}).
In addition, the output feature map (i.e., the input feature map of the next layer) of the temporal convolution, $\mathbf{Y_{1d}} \in \mathbb{R}^{t \times 1\times c^{\prime}}$, is smaller than that of a 2D convolution, $\mathbf{Y_{2d}} \in \mathbb{R}^{t \times f\times c}$.
The decrease in feature map size leads to a dramatic reduction of the computational burden and footprint in the following layers, which is key to implementing fast KWS.

\subsection{TC-ResNet Architecture} \label{subsection:architecture}

\begin{figure}[t]
    \includegraphics[width=0.47\textwidth]{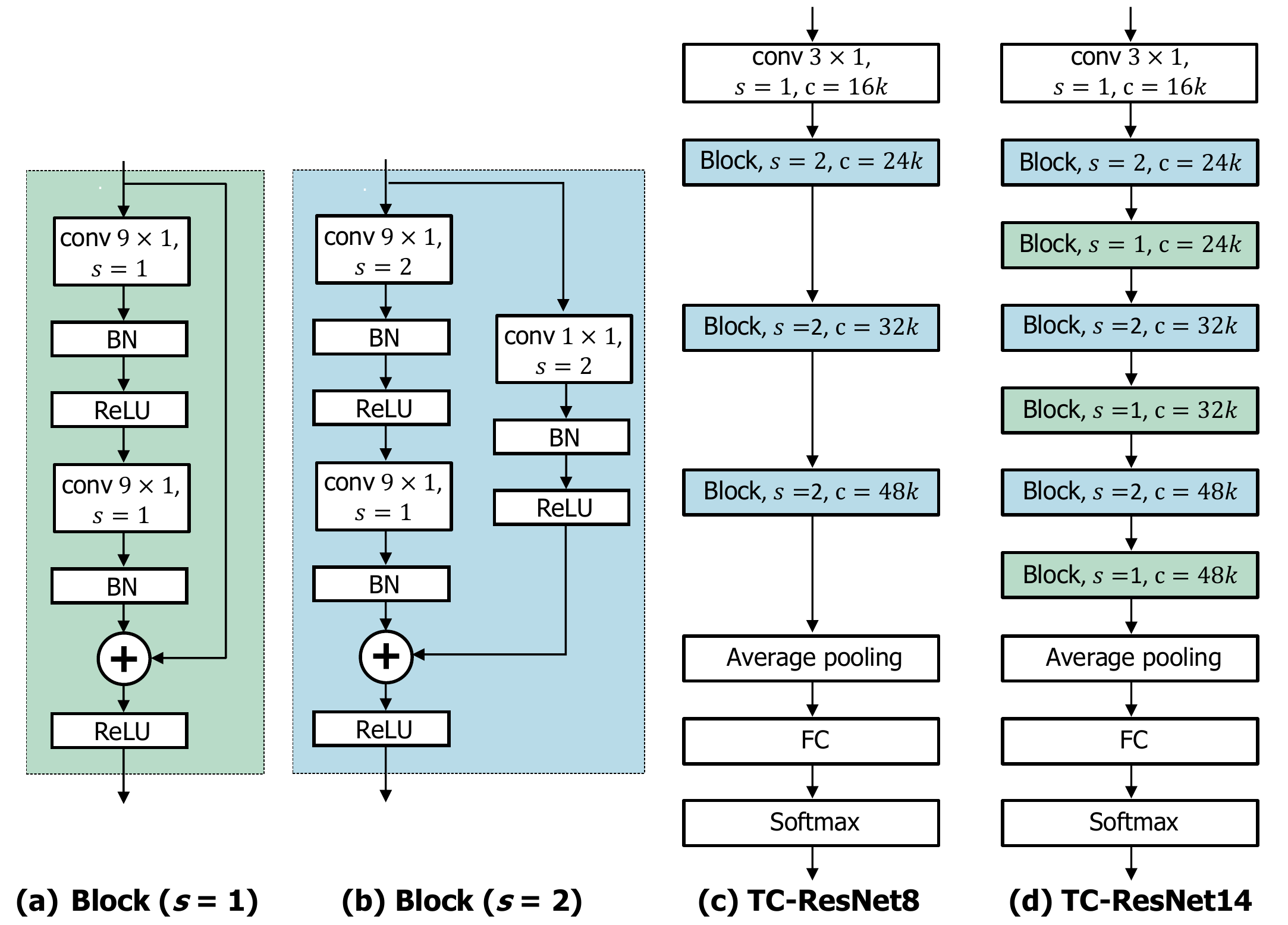}
    \caption{The building block (denoted Block) of TC-ResNet when (a) stride = 1 and (b) stride = 2.
    (c) Architecture for TC-ResNet8 and (d) TC-ResNet14.
    Each of them utilizes ResNet8 and ResNet14 as the backbone-CNN, respectively.
    BN and FC denote batch normalization and fully connected layer.
    Note that `s', `c', and `k' indicates stride, channel size, and width multiplier, respectively.
    }
    \vspace{-0.2cm}
    \label{fig:model}
\end{figure}

We adopt ResNet~\cite{he-cvpr-2016-resnet}, one of the most widely used CNN architectures, but utilize $m \times 1$ kernels ($m = 3$ for the first layer and $m = 9$ for the other layers) rather than $3 \times 3$ kernels (Figure~\ref{fig:model}).
None of the convolution layers and fully connected layers have biases, and each batch normalization layer~\cite{ioffe-arxiv-2015-batchnormalization} has trainable parameters for scaling and shifting.
The identity shortcuts can be directly used when the input and the output have matching dimensions (Figure~\ref{fig:model}a), otherwise, we use an extra \emph{conv-BN-ReLU} to match the dimensions (Figure~\ref{fig:model}b).
Tang and Lin~\cite{tang-icassp-2018-residual} also adopted the residual network, but they did not employ a temporal convolution and used a conventional $3 \times 3$ kernel.
In addition, they replaced strided convolutions with dilated convolutions of stride one.
Instead, we employ temporal convolutions to increase the effective receptive field and follow the original ResNet implementation for other layers by adopting strided convolutions and excluding dilated convolutions.

We select \textbf{\emph{TC-ResNet8}} (Figure~\ref{fig:model}c), which has three residual blocks and $\{16, 24, 32, 48\}$ channels for each layer including the first convolution layer, as our base model.
\textbf{\emph{TC-ResNet14}} (Figure~\ref{fig:model}d) expands the network by incorporating twice as much residual blocks compared to \emph{TC-ResNet8}.

We introduce width multiplier~\cite{howard-arxiv-2017-mobilenet} (\emph{k} in Figure~\ref{fig:model}c and Figure~\ref{fig:model}d) to increase (or decrease) the number of channels at each layer, thereby achieving flexibility in selecting the right capacity model for given constraints.
For example, in \emph{TC-ResNet8}, a width multiplier of $1.5$ expands the model to have $\{24, 36, 48, 72\}$ number of channels respectively.
We denote such a model by appending a multiplier suffix such as \textbf{\emph{TC-ResNet8-1.5}}.
\textbf{\emph{TC-ResNet14-1.5}} is created in the same manner.

\section{Experimental Framework}\label{section:implementation}

\subsection{Experimental Setup} \label{subsection:experimentalsetup}
\textbf{Dataset.}
We evaluated the proposed models and baselines~\cite{sainath-interspeech-2015-cnn, tang-icassp-2018-residual, zhang-arxiv-2017-helloedge} using \emph{Google Speech Commands Dataset}~\cite{googlespeechcommandsv1}.
The dataset contains 64,727 one-second-long utterance files which are recorded and labeled with one of 30 target categories.
Following Google's implementation~\cite{googlespeechcommandsv1}, we distinguish 12 classes: \emph{``yes'', ``no'', ``up'', ``down'', ``left'', ``right'', ``on'', ``off'', ``stop'', ``go'', silence}, and \emph{unknown}.
Using SHA-1 hashed name of the audio files, we split the dataset into training, validation, and test sets, with 80\% training, 10\% validation, and 10\% test, respectively.

\textbf{Data augmentation and preprocessing.}
We followed Google's preprocessing procedures which apply random shift and noise injection to training data.
First, in order to generate background noise, we randomly sample and crop background noises provided in the dataset, and multiply it with a random coefficient sampled from uniform distribution, $U(0, 0.1)$.
The audio file is decoded to a float tensor and shifted by $s$ seconds with zero padding, where $s$ is sampled from $U(-0.1, 0.1)$.
Then, it is blended with the background noise.
The raw audio is decomposed into a sequence of frames following the settings of the previous study~\cite{tang-icassp-2018-residual} where the window length is 30 \emph{ms} and the stride is 10 \emph{ms} for feature extraction.
We use 40 MFCC features for each frame and stack them over time-axis.

\textbf{Training.}
We trained and evaluated the models using TensorFlow~\cite{abadi-osdi-2016-TensorFlow}. 
We use a weight decay of 0.001 and dropout with a probability of 0.5 to alleviate overfitting.
Stochastic gradient descent is used with a momentum of 0.9 on a mini-batch of 100 samples.
Models are trained from scratch for 30k iterations.
Learning rate starts at 0.1 and is divided by 10 at every 10k iterations.
We employ early stopping~\cite{prechelt-springer-1998-earlystop} with the validation split.

\textbf{Evaluation.}
We use \emph{accuracy} as the main metric to evaluate how well the model performs.
We trained each model 15 times and report its average performance.
\emph{Receiver operating characteristic (ROC) curves}, of which the $x$-axis is the false alarm rate and the $y$-axis is the false reject rate, are plotted to compare different models.
To extend the ROC curve to multi-classes, we perform micro-averaging over multiple classes per experiment, then vertically average them over the experiments for the final plot.

We report the number of operations and parameters which faithfully reflect the real-world environment for mobile deployment.
Unlike previous works which only reported the numbers for part of the computation such as the number of multiply operations~\cite{tang-icassp-2018-residual} or the number of multiplications and additions only in the matrix-multiplication operations~\cite{zhang-arxiv-2017-helloedge}, we include \emph{FLOPs}~\cite{arik-arxiv-2018-tensorflowprofiler}, computed by TensorFlow profiling tool~\cite{tensorflow-profile}, and the number of \emph{all} parameters instead of only trainable parameters reported by previous studies~\cite{tang-icassp-2018-residual}.

Inference speed can be estimated by FLOPs but it is well known that FLOPs are not always proportional to speed.
Therefore, we also measure \emph{inference time} on a mobile device using the TensorFlow Lite Android benchmark tool~\cite{tflite-model-benchmark-tool}.
We measured inference time on a Google Pixel 1 and forced the model to be executed on a single little core in order to emulate the always-on nature of KWS.
The benchmark program measures the inference time 50 times for each model and reports the average.
Note that the inference time is measured from the first layer of models that receives MFCC as input to focus on the performance of the model itself.

\subsection{Baseline Implementations}
We carefully selected baselines and verified advantages of the proposed models in terms of accuracy, the number of parameters, FLOPs, and inference time on mobile devices.
Below are the baseline models:

\begin{itemize}
\item \textbf{\emph{CNN-1}} and \textbf{\emph{CNN-2}}~\cite{sainath-interspeech-2015-cnn}.
We followed the implementations of~\cite{zhang-arxiv-2017-helloedge} where window size is 40 \emph{ms} and the stride is 20 \emph{ms} using 40 MFCC features.
\emph{CNN-1} and \emph{CNN-2} represent \emph{cnn-trad-fpool3} and \emph{cnn-one-fstride4} in \cite{sainath-interspeech-2015-cnn}, respectively.
\item \textbf{\emph{DS-CNN-S}}, \textbf{\emph{DS-CNN-M}}, and \textbf{\emph{DS-CNN-L}}~\cite{zhang-arxiv-2017-helloedge}. 
\emph{DS-CNN} utilizes depthwise convolutions. 
It aims to achieve the best accuracy when memory and computation resources are constrained.
We followed the implementation of~\cite{zhang-arxiv-2017-helloedge} which utilizes 40 \emph{ms} window size with 20 \emph{ms} stride and only uses 10 MFCCs to reduce the number of operations.
\emph{DS-CNN-S}, \emph{DS-CNN-M}, and \emph{DS-CNN-L} represent small-, medium-, and large-size model, respectively.
\item \textbf{\emph{Res8}}, \textbf{\emph{Res8-Narrow}}, \textbf{\emph{Res15}}, and \textbf{\emph{Res15-Narrow}}~\cite{tang-icassp-2018-residual}.
\emph{Res}-variants employ a residual architecture for keyword spotting.
The number following \emph{Res} (e.g., 8 and 15) denotes the number of layers and the \emph{-Narrow} suffix represents that the number of channels is reduced.
\emph{Res15} has shown the best accuracy with Google Speech Commands Dataset among the KWS studies which are based on CNNs.
The window size is 30 \emph{ms}, the stride is 10 \emph{ms}, and MFCC feature size is 40.
\end{itemize}
We release our end-to-end pipeline codebase for training, evaluating, and benchmarking the baseline models and together with the proposed models.
It consists of TensorFlow implementation of models, scripts to convert the models into the TensorFlow Lite models that can run on mobile devices, and the pre-built TensorFlow Lite Android benchmark tool.

\section{Experimental Results}\label{section:results}

\subsection{Google Speech Command Dataset}

\begin{table}[t]
    \begin{tabular}{lcccc}
        \toprule
        \multirow{2}{*}{Model} & Acc. & Time & FLOPs & Params \\
                              & (\%) & (\emph{ms})     &       & \\
        \midrule
        CNN-1                   & 90.7$^{\star}$ & 32  & 76.1M  & 524K \\
        CNN-2                   & 84.6$^{\star}$ & \textbf{1.2}  & 1.5M  & 148K \\
        DS-CNN-S                & 94.4$^{\star}$ & 1.6   & 5.4M  & 24K \\
        DS-CNN-M                & 94.9$^{\star}$ & 5.2   & 19.8M & 140K \\
        DS-CNN-L                & 95.4$^{\star}$ & 16.8  & 56.9M & 420K \\        
        Res8-Narrow             & 90.1$^{\star}$ & 47    & 143.2M  & 20K \\
        Res8                    & 94.1$^{\star}$ & 174   & 795.3M  & 111K \\
        Res15-Narrow            & 94.0$^{\star}$ & 107   & 348.7M  & 43K \\
        Res15                   & \textbf{95.8}$^{\star}$ & 424   & 1950.0M  & 239K \\
        \midrule
        TC-ResNet8                   & 96.1 & \textbf{1.1}   & 3.0M  & 66K \\
        TC-ResNet8-1.5               & 96.2 & 2.8   & 6.6M  & 145K \\
        TC-ResNet14                  & 96.2 & 2.5   & 6.1M  & 137K \\
        TC-ResNet14-1.5              & \textbf{96.6} & 5.7   & 13.4M & 305K \\
        \bottomrule
    \end{tabular}
    \caption{Comparison of the baseline models and the proposed models. 
    The numbers marked with $\star$ are taken from the paper.
    The best result (accuracy and latency) among different approaches are displayed in bold.
    }
    \vspace{-0.3cm}
    \label{tab:res}
\end{table}

\begin{figure}[t]
    \includegraphics[width=0.45\textwidth]{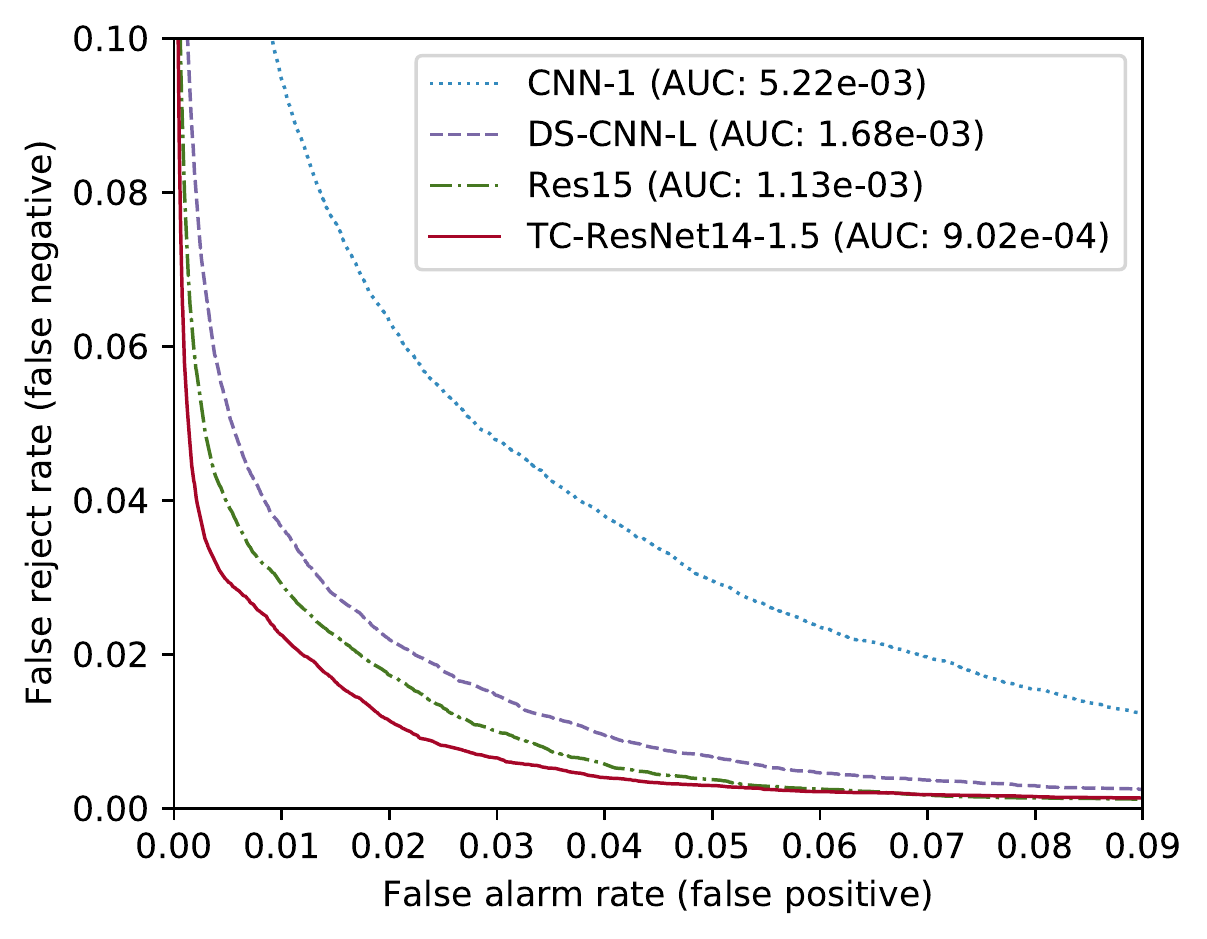}
    \caption{ROC curves for selected models with corresponding values of AUC.}
    \vspace{-0.2cm}
    \label{fig:roc}
\end{figure}

Table~\ref{tab:res} shows the experimental results.
Utilizing advantages of temporal convolutions, we improve the inference time measured on mobile device dramatically while achieving better accuracy compared to the baseline KWS models.
\textit{TC-ResNet8} achieves 29x speedup while improving 5.4\%p in accuracy compared to \textit{CNN-1}, and improves 11.5\%p in accuracy while maintaining a comparable latency to \textit{CNN-2}.
Since \emph{DS-CNN} is designed for the resource-constrained environment, it shows better accuracy compared to the naive CNN models without using large number of computations.
However, \emph{TC-ResNet8} achieves 1.5x / 4.7x / 15.3x speedup, and improves 1.7\%p / 1.2\%p / 0.7\%p accuracy compared to \emph{DS-CNN-S} / \emph{DS-CNN-M} / \emph{DS-CNN-L}, respectively.
In addition, the proposed models show better accuracy and speed compared to \emph{Res} which shows the best accuracy among baselines. 
\emph{TC-ResNet8} achieves 385x speedup while improving 0.3\%p accuracy compared to deep and complex \emph{Res} baseline, \emph{Res15}.
Compared to a slimmer \emph{Res} baseline, \emph{Res8-Narrow}, proposed \emph{TC-ResNet8} achieves 43x speedup while improving 6\%p accuracy.
Note that our wider and deeper models (e.g., \textit{TC-ResNet8-1.5}, \textit{TC-ResNet14}, and \textit{TC-ResNet14-1.5}) achieve better accuracy at the expense of inference speed.

We also plot the ROC curves of models which depict the best accuracy among their variants: \emph{CNN-1}, \emph{DS-CNN-L}, \emph{Res15}, and \emph{TC-ResNet14-1.5}.
As presented in Figure~\ref{fig:roc}, \emph{TC-ResNet14-1.5} is less likely to miss target keywords compared to other baselines assuming that the number of incorrectly detected keywords is the same.
The small area under the curve (AUC) means that the model would miss fewer target keywords on average for various false alarm rates.
\emph{TC-ResNet14-1.5} shows the smallest AUC, which is critical for good user experience with KWS system.

\subsection{Impact of Temporal Convolution}

\begin{table}[h]
    \begin{tabular}{lcccc}
        \toprule
        \multirow{2}{*}{Model} & Acc. & Time & FLOPs & Params \\
                              & (\%) & (\emph{ms})     &       & \\
        \midrule
        2D-ResNet8              & 96.1 & 10.1  & 35.8M & 66K \\
        2D-ResNet8-Pool         & 94.9 & 3.5   & 4.0M  & 66K \\
        \bottomrule
    \end{tabular}
    \caption{Comparison of TC-ResNet variants, 2D-ResNet8 and 2D-ResNet8-Pool, which utilize 2D convolutions while retaining the architecture and the number of parameters of TC-ResNet8.
    }
    \vspace{-0.6cm}
    \label{tab:impact-1d}
\end{table}

We demonstrate that the proposed method could effectively improve both accuracy and inference speed compared to the baseline models which treat the feature map as a 2D image.
We further explore the impact of the temporal convolution by comparing variants of \emph{TC-ResNet8}, named \emph{2D-ResNet8} and \emph{2D-ResNet8-Pool}, which adopt a similar network architecture and the number of parameters but utilize 2D convolutions.

We designed \emph{2D-ResNet8}, whose architecture is identical to \emph{TC-ResNet8} except for the use of $3 \times 3$ 2D convolutions.
\emph{2D-ResNet8} (in Table~\ref{tab:impact-1d}) shows comparable accuracy, but is 9.2x slower compared to \emph{TC-ResNet8} (in Table~\ref{tab:res}).
\emph{TC-ResNet8-1.5} is able to surpass \emph{2D-ResNet8} while using less computational resources.

We also demonstrate the use of temporal convolution is superior to other methods of reducing the number of operations in CNNs such as applying a pooling layer.
In order to reduce the number of operations while minimizing the accuracy loss, \emph{CNN-1}, \emph{Res8}, and \emph{Res8-Narrow} adopt average pooling at an early stage, specifically, right after the first convolution layer.
We inserted an average pooling layer, where both the window size and the stride are set to 4, after the first convolution layer of \emph{2D-ResNet8}, and named it \emph{2D-ResNet8-Pool}.
\emph{2D-ResNet8-Pool} improves inference time with the same number of parameters, however, it loses 1.2\%p accuracy and is still 3.2x slower compared to \emph{TC-ResNet8}.

\section{Related Works} \label{section:relatedwork}

Recently, there has been a wide adoption of CNNs in KWS.
Sainath \emph{et~al}.~\cite{sainath-interspeech-2015-cnn} proposed small-footprint CNN models for KWS.
Zhang \emph{et~al}.~\cite{zhang-arxiv-2017-helloedge} searched and evaluated proper neural network architectures within memory and computation constraints.
Tang and Lin~\cite{tang-icassp-2018-residual} exploited residual architecture and dilated convolutions to achieve further improvement in accuracy while preserving compact models.
In previous studies~\cite{sainath-interspeech-2015-cnn, zhang-arxiv-2017-helloedge, tang-icassp-2018-residual}, it has been common to use 2D convolutions for inputs with time-frequency representations.
However, there has been an increase in the use of 1D convolutions in acoustics and speech domain~\cite{lim-dcase-2017-rare, choi-icassp-2017-music}.
Unlike previous studies~\cite{lim-dcase-2017-rare, choi-icassp-2017-music} our work applies 1D convolution along the temporal axis of time-frequency representations instead of convolving along the frequency axis or processing raw audio signals.



\section{Conclusion} \label{section:conclusion}
In this investigation, we aimed to implement \emph{fast} and \emph{accurate} models for real-time KWS on mobile devices.
We measured inference speed on the mobile device, Google Pixel 1, and provided quantitative analysis of conventional convolution-based KWS models and our models utilizing temporal convolutions.
Our proposed model achieved 385x speedup while improving 0.3\%p accuracy compared to the state-of-the-art model.
Through ablation study, we demonstrated that temporal convolution is indeed responsible for the dramatic speedup while improving the accuracy of the model.
Further studies analyzing the efficacy of temporal convolutions for a diverse set of network architectures would be worthwhile.

\bibliographystyle{IEEEtran}
\bibliography{mybib}

\end{document}